\documentclass[sigconf]{acmart}
\AtBeginDocument{%
  }

\copyrightyear{2025} 
\acmYear{2025} 
\setcopyright{acmlicensed}\acmConference[DIS '25 Companion]{Designing Interactive Systems Conference}{July 5--9, 2025}{Funchal, Portugal}
\acmBooktitle{Designing Interactive Systems Conference (DIS '25 Companion), July 5--9, 2025, Funchal, Portugal}
\acmDOI{10.1145/3715668.3736337}
\acmISBN{979-8-4007-1486-3/2025/07}

\begin{document}

\title{“Gaze and Glow”: Exploring Editing Processes on Social Media through Interactive Exhibition}

\author{Yang Hong}
\authornotemark[1]
\affiliation{%
  \institution{National Yang Ming Chiao Tung University}
  \city{Hsinchu}
  \country{Taiwan}}
\email{hongyang.hk12@nycu.edu.tw}

\author{Jie-Yi Feng}
\affiliation{%
  \institution{OneDegree HK Team}
  \city{Taipei}
  \country{Taiwan}}
\email{nicole.feng@onedegree.hk}

\author{Yi-Chun Yao}
\affiliation{%
  \institution{National Yang Ming Chiao Tung University}
  \city{Hsinchu}
  \country{Taiwan}}
\email{yeeyao.hk08@nycu.edu.tw}

\author{I-Hsuan Cho}
\affiliation{%
 \institution{Tenten Creative}
 \city{Taipei}
 \country{Taiwan}}
\email{chosali2000@gmail.com}

\author{Yu-Ting Lin}
\affiliation{%
  \institution{National Yang-Ming Chiao Tung University}
  \city{Taipei}
  \country{Taiwan}}
\email{yutinglinw@gmail.com}

\author{Ying-Yu Chen}
\affiliation{%
  \institution{National Yang Ming Chiao Tung University}
  \city{Hsinchu}
  \country{Taiwan}}
\email{yingyuchen@nycu.edu.tw}
\renewcommand{\shortauthors}{Yang et al.}

\begin{abstract}
We present "Gaze and Glow," an interactive installation that reveals the often-invisible efforts of social media editing. Through narrative personas, experimental videos, and sensor-based interactions, the installation explores how audience attention shapes users’ editing practices and emotional experiences. Deployed in a two-month public exhibition, Gaze and Glow engaged viewers and elicited responses. Reflexive thematic analysis of audience feedback highlights how making editing visible prompts new reflections on authenticity, agency, and performativity. We discuss implications for designing interactive systems that support selective memory, user-controlled visibility, and critical engagement with everyday digital self-presentation.
\end{abstract}


\begin{teaserfigure}
  \includegraphics[width=\textwidth]{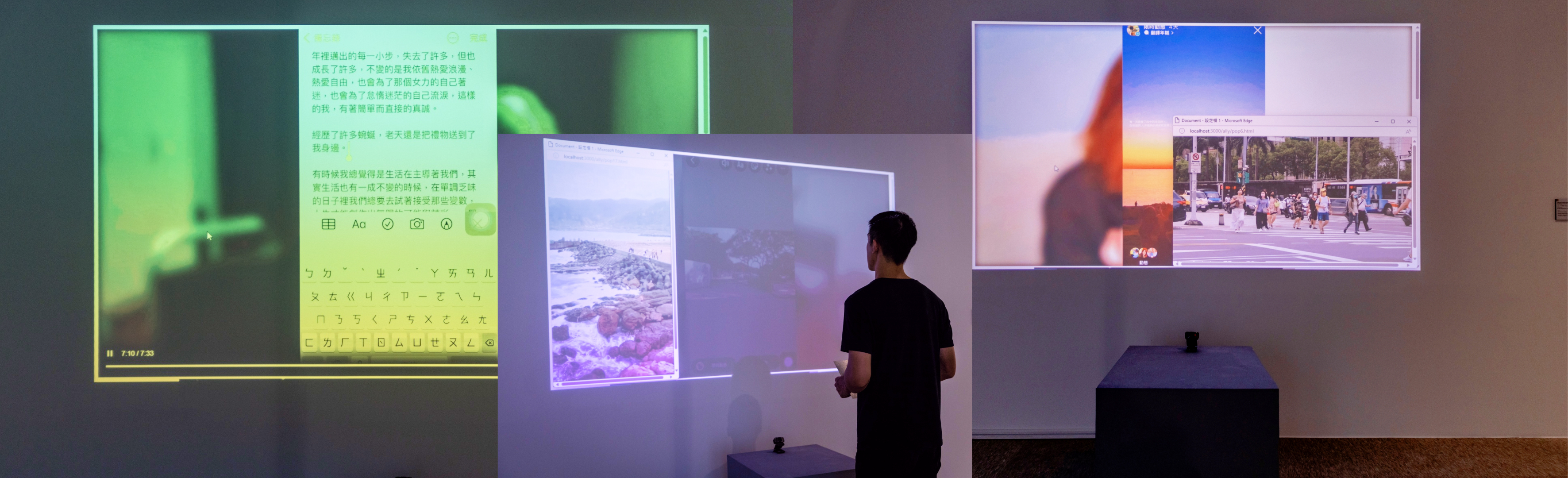}
  \caption{The interactive installation "Gaze and Glow" in a public exhibition}
  \Description{key picture}
  \label{fig:images/key.jpg}
\end{teaserfigure}


\maketitle

\section{Introduction}
The editing process on social media before publishing is often regarded as a private and ephemeral activity, or a “deceptive” deviation from authenticity, such as in the case of modifications in dating profiles to fabricate online identities \cite{van_gelder_strange_1995} or users' meticulous revisions to ensure curated positive and ideal self-images \cite{reddy_teaspoon_2024, mills_selfie_2018}. Previous research in HCI and social media studies has examined users' engagements in selfie retouching or text polishing before content publishing \cite{tiggemann_uploading_2020, ying_retouchingffhq_2023, taber_finsta_2020, trieu_private_2020}. However, these studies have primarily focused on the outcomes of the editing behavior, such as concerns on online authenticity \cite{ying_retouchingffhq_2023, taber_finsta_2020}, or its cognitive influence on perceived attractiveness \cite{ozimek_how_2023} and body dissatisfaction \cite{tiggemann_uploading_2020}; while paying less attention to the socially situated considerations and emotional efforts behind the unfolded editing processes.

In this work, we investigate the potential to expose and amplify the often untraceable editing processes behind social media content creation by staging them in a public, interactive exhibition. We conceptualize these processes not as isolated or purely technical steps, but as thoughtful, dynamic, and socially situated experiences. Drawing inspiration from prior research in digital curation and critical artistic practices on Instagram \cite{ulman_excellences_2018}, we develop narrative personas and reconstruct their editing practices prior to publishing Stories and Posts on Instagram. By making these otherwise invisible practices visible within a public exhibition space, we examine how the act of revealing the behind-the-scenes editing work of social media users may reshape public understanding of self-presentation, and analyze how audiences respond to these disclosures.

To achieve this, we employed a combination of cultural probes and semi-structured interviews to develop \textit{Gaze and Glow}, an interactive installation that reveals the editing routines and emotions of three narrative personas. The installation integrates experimental video narratives with ultrasonic distance sensors, which not only depict the nuanced contexts and internal dialogues during the editing process but also allow the audience’s physical proximity to influence the unfolding of the narratives. In doing so, the installation foregrounds how social media editing is both a personal and performative act, shaped by imagined or real audiences \cite{bernstein_quantifying_2013}.

We deployed Gaze and Glow in a two-month public exhibition at a city museum with free admission, during which we collected 54 written responses from nearly one hundred visitors who interacted with the installation. These responses included reflections on the visibility of editing practices, self-recognition in the personas, and thoughts on social media culture. We analyzed the audience feedback to examine how viewers engaged with the installation, and how the exposure of previously invisible editing practices prompted new interpretations of authenticity, agency, and performativity in social media use.

\section{Designing “Gaze and Glow”}

\subsection{Cultural Probe and Semi-structured Interview}

We first conducted a two-week cultural probe study \cite{gaver_design_1999} to explore participants' editing experiences during their daily use of Instagram. Cultural probe packages were mailed to participants and included blank profile cards for illustrating their overall editing styles on different Instagram accounts, with instructions for recording their editing processes through online diaries over the two weeks. They were supposed to upload the duration and steps applied to editing text, images and videos before publishing on Instagram, and their motivations during the processes. Participants were encouraged to share screenshots, recordings, and relevant text descriptions. Then, we conducted follow-up interviews to learn more about the data collected through the cultural probe. Participants provided additional details and perspectives on their practices, feelings, motivations, and considerations during the interviews.

\begin{figure}[htbp]
  \centering
  \includegraphics[width=\linewidth]{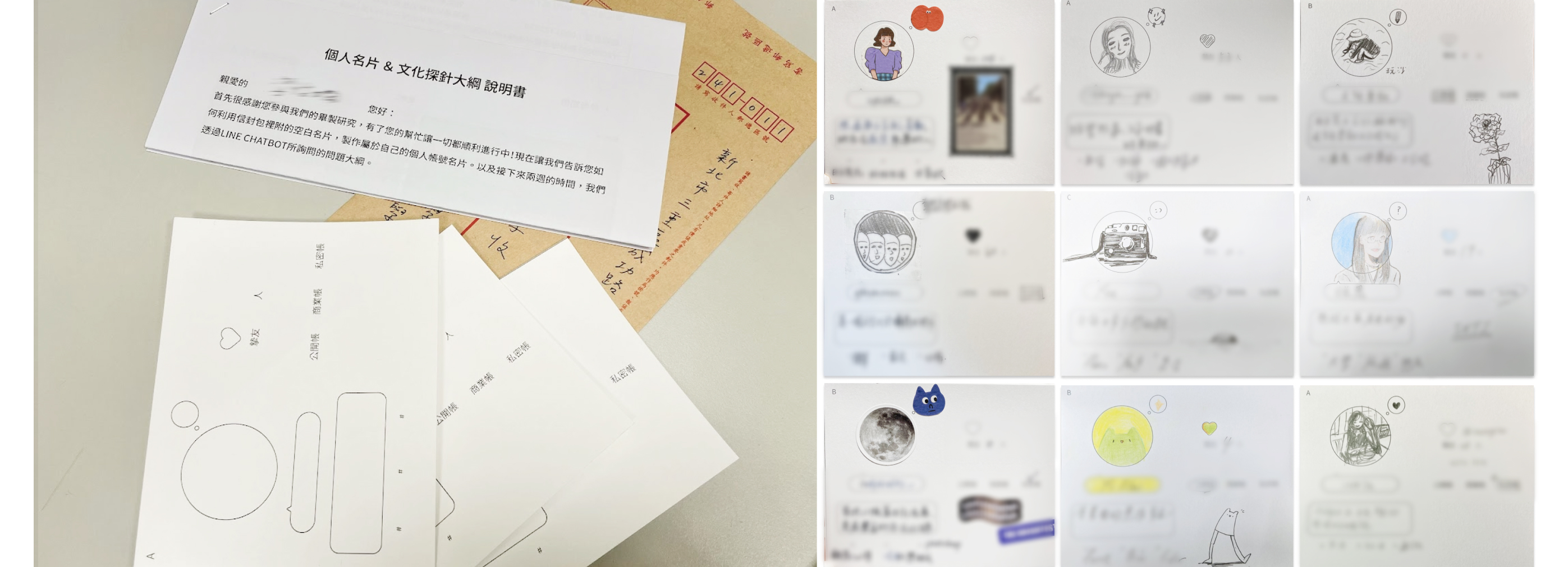}
  \caption{the cultural probe packages, instructions (in Chinese) and examples of participants’ profile cards}
  \Description{the cultural probe package}
\end{figure}

\subsection{Designing “Gaze and Glow”: From Persona to Experimental Videos}

We analyzed qualitative data using Affinity Diagramming \cite{iba_philosophy_2017}, which helped us identify key motivations, strategies, and contextual factors in participants’ editing practices. This analysis guided the development of three narrative personas, each embodied a range of editing behaviors, including visual modifications (e.g., retouching, filter selection, collages, layout adjustments) and textual revisions (e.g., refining language, formatting, supplementing content). Editing practices were revealed to be highly dynamic, context-dependent, and socially influenced, shaped by content type (e.g., ephemeral Stories vs. permanent Posts), account type (e.g., main vs. secondary accounts), and the perceived significance of the content (e.g., milestone events such as birthdays or graduations often prompted more extensive revisions). Rather than a linear or solitary task, editing was experienced as a recursive and socially entangled process, shaped by interactions with specific followers as well as a broader, imagined audience. Participants described editing as a way to anticipate, manage, and respond to audience perception.

To contextualize the editing behaviors portrayed in the installation, we provided demographic and motivational details for each of the three narrative personas. Lily Ho, a 20-year-old college student from Taipei, regards editing as a means to ensure that everyone appears their best in her posts. Her approach is highly visual and meticulous, particularly for milestone events, with a stronger emphasis on image composition than on textual content. Ally Chen, a 28-year-old designer and photographer, approaches editing as a poetic and aesthetic act of curation. She places high value on both visual and textual refinement, aiming to craft emotionally expressive and stylistically cohesive posts. Toby Lee, a 24-year-old freelance creative, uses filters to enhance the realism of captured moments. Although his editing is less intensive, it remains intentional, and he maintains that such edits do not compromise the authenticity of the content. These personas represent a range of editing motivations spanning impression management, artistic expression, and the pursuit of authenticity.

\begin{figure}[htbp]
  \centering
  \includegraphics[width= 0.65\linewidth]{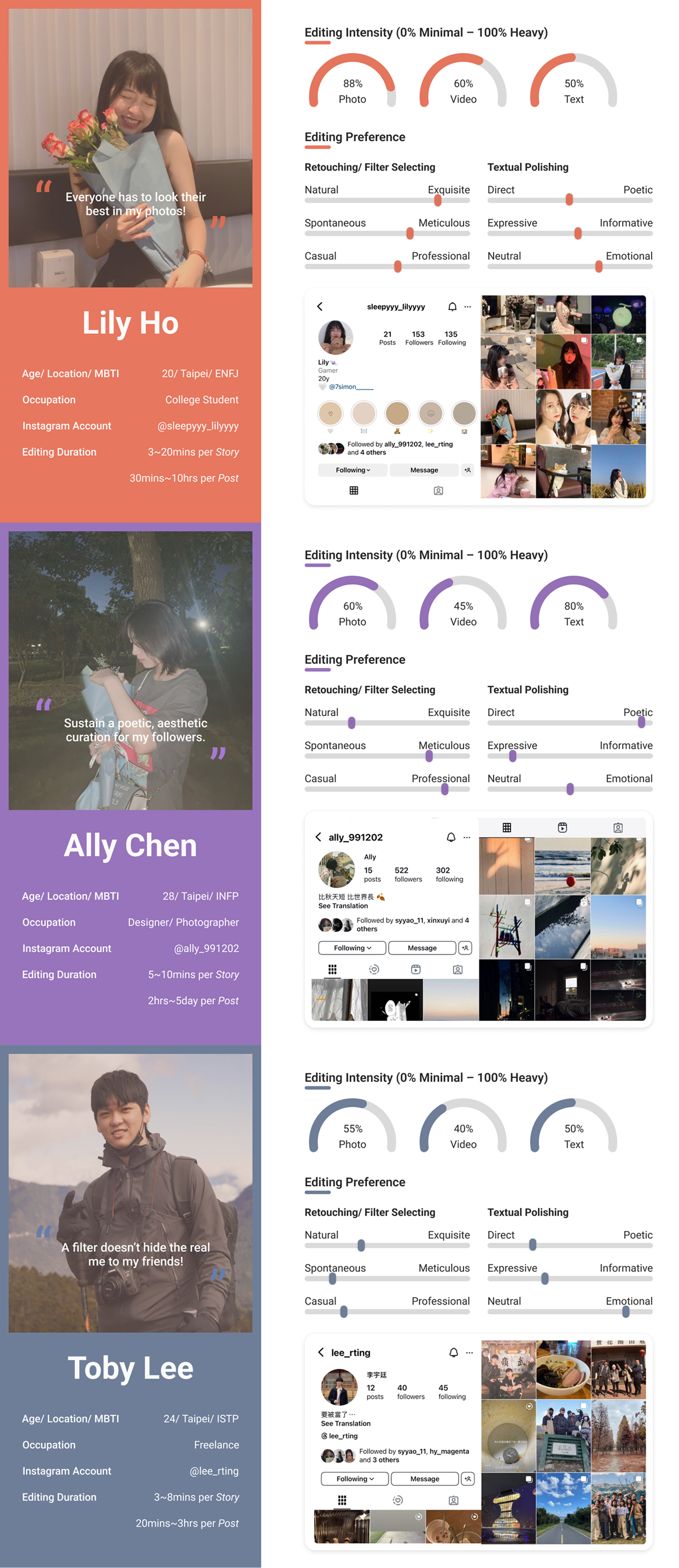}
  \caption{Three narrative personas constructed from cultural probe and interview data with varied editing preferences and practices.}
  \Description{personas}
\end{figure}

Building on these insights, we created a series of experimental videos to materialize and externalize each persona’s editing process. Inspired by stories from participants who spent significant time crafting and revising their birthday posts, we focused on the “birthday” scenario, a moment often associated with heightened self-presentation. The videos adopted a Vlog-style format interwoven with screen recordings of the editing interface, depicting the nonlinear, fluid, and often fragmented nature of the editing experience. The videos featured a sequence of daily activities punctuated by moments of focused editing, mirroring how participants engaged with content intermittently throughout the day. To complement the videos, we also created Instagram accounts for each persona, which showcased the final posts and Stories referenced in the videos, extending the narrative into a familiar social media format.

\begin{figure}[htbp]
  \centering
  \includegraphics[width=\linewidth]{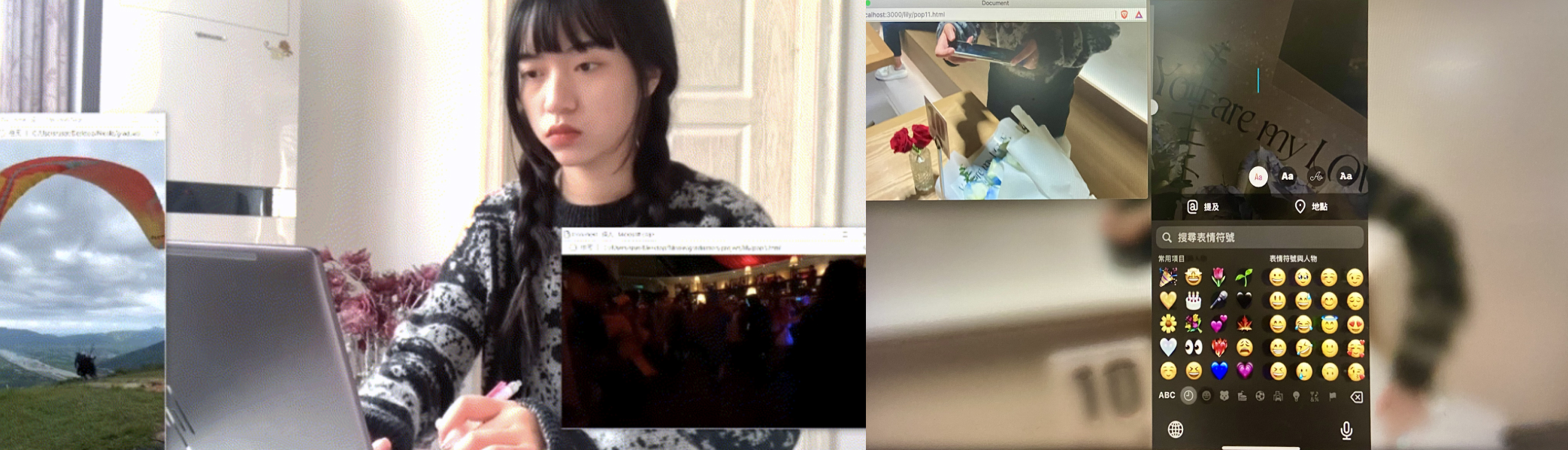}
  \caption{The experimental Vlogs show the editing process offline and online, including personas' state of life before editing the article, personas' mobile screens as they write text and edit pictures, and pop-ups from their process of selecting material.}
  \Description{video}
\end{figure}

To make the installation interactive and responsive to audience presence, we drew on Gaver et al.’s (2003) notion of “ambiguity of relationship” to encourage open-ended interpretation rather than prescriptive messaging \cite{gaver_ambiguity_2003}. These reflections led us to explore how viewer proximity and presence might be incorporated into the installation to simulate the affective impact of perceived audience scrutiny. We used ultrasonic distance sensors to detect when a viewer approached the screen. As the viewer moved closer, the screen’s brightness and saturation gradually decreased, metaphorically representing the increased pressure the persona felt under close attention. In addition, we used face-api.js and embedded cameras to calculate the number of viewers in front of the screen. As more people gathered, the system triggered a higher frequency of pop-up messages visualizing the links between audience size and perceived stress during the editing process. We prototyped and iterated on the interaction design prior to the exhibition.

\begin{figure}[htbp]
  \centering
  \includegraphics[width=\linewidth]{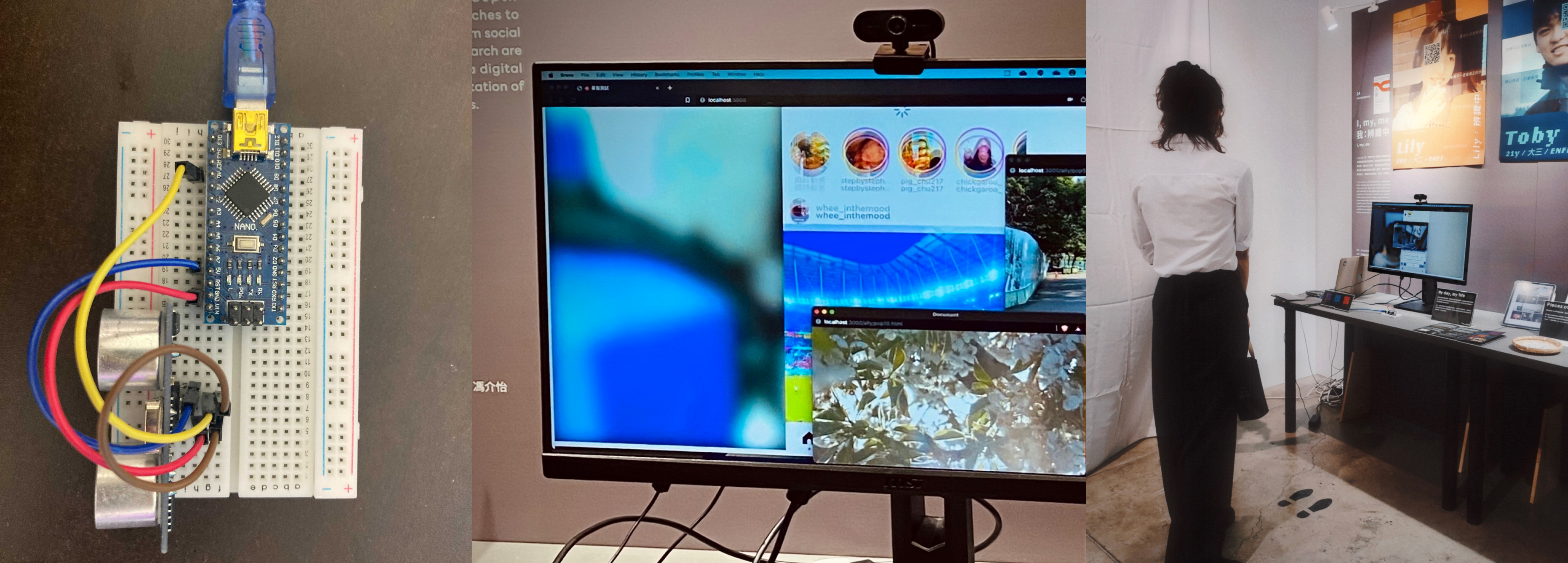}
  \caption{Ultrasonic distance sensor, Arduino Nano, and a prototype version of "Gaze and Glow"}
  \Description{prototype}
\end{figure}

\section{Deploying Gaze and Glow: Viewer Feedback}

The installation was deployed over a two-month period in a public exhibition space, during which we gathered 54 responses from viewers aged 17 to 51 through open-ended online questionnaires and informal on-site interviews. We analyzed the responses using reflexive thematic analysis \cite{braun_thematic_2020} to understand how viewers engaged with the installation and how it shaped their reflections on social media editing practices.

\subsubsection*{Making Editing Public: Shared Exploration of Interaction}
Many viewers initially expressed uncertainty about how to interact with Gaze and Glow. Upon noticing the dimming effect as they moved closer to the screen, they began to experiment — walking back and forth to test the system’s response. Several participants engaged their friends in these interactions, sparking conversations about how the system responded to proximity and group presence. This playful exploration transformed editing, typically a solitary and private activity, into a shared and open experience. By externalizing the process of editing through embodied interaction, the installation invited viewers to co-investigate how visibility and attention shape editing behaviors. The installation thus served not only as a representation of editing but also as a catalyst for dialogue and collective interpretation.

\subsubsection*{Mixed Feelings: Recognition and Discomfort}

Engaging with the installation prompted viewers to reflect on their own editing routines — practices that often go unacknowledged. For some, the amplified visibility of the editing process evoked discomfort. Five participants described feelings of embarrassment or vulnerability in seeing the private, often trivial, details of their editing practices rendered visible in a public space. One viewer noted:
	\textit{“I felt shameful to think about all the trivial details of editing being exposed. When I moved closer to the screen and saw it dim, I felt relieved.” (23, F)}
Conversely, other viewers experienced a sense of validation and recognition. Twelve participants reported that they had never consciously considered their editing processes before, and were surprised by how much efforts and emotion went into them. For these viewers, the installation legitimized the invisible effort they invested in curating posts. As one participant reflected:
	\textit{“I think the emotional struggles during the editing process deserve to be seen.” (26, F)}
This contrast in responses highlights the dual nature of editing as both empowering and anxiety, a private activity that can feel exposed or affirmed when made visible.

\subsubsection*{Rethinking Authenticity Through Selective Disclosure}
Some viewers suggested that exposing parts of the editing process could help balance the tension between curated online content and offline authenticity. Rather than undermining self-presentation, selective disclosure was seen as a way to enrich it. One participant noted:
	\textit{“It reveals more dimensions of her (one of the personas) personality, not just the polished final image.” (29, F)}
Others shared personal practices of revealing their editing process as a way to maintain a sense of honesty. For example:
	\textit{“I sometimes tell friends in the comments that my good selfie is ‘heavily edited’.” (21, F)}
These responses suggest that authenticity is not necessarily the absence of editing, but rather the willingness to acknowledge or selectively disclose the efforts behind the polished image. By making editing visible in a nuanced way, Gaze and Glow encouraged viewers to reflect on how realness can coexist with curation.

\begin{figure}[htbp]
  \centering
  \includegraphics[width=\linewidth]{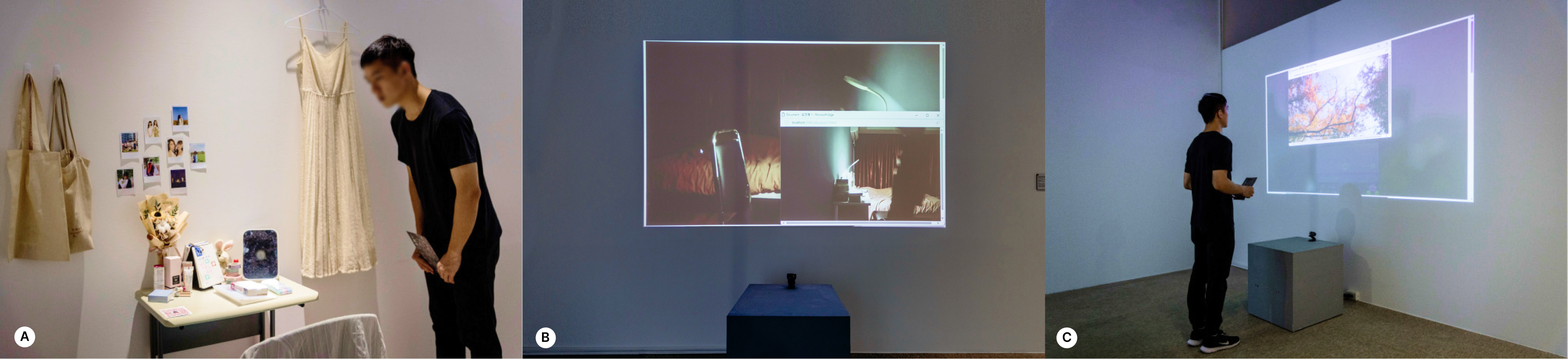}
  \caption{To show the offline contexts of editing, we designed a small corner in the exhibition containing stuff belonging to three personas and appearing in their posts. (A) A viewer examined the stuff of one of the personas. (B) The installation was playing the Vlog of personas’ editing process. (C) A viewer interacted with the installation.}
  \Description{video}
\end{figure}

\section{Discussion}
Gaze and Glow served both as a reflective tool and a design provocation, making the invisible efforts of social media editing tangible, visible, and interactive. In this section, we reflect on our design process and propose directions for how platforms might reconsider editing as a meaningful site of memory and selective visibility, rather than a transient, disposable activity.

\subsubsection*{Designing to Reveal the Editing Process}
The design and deployment of Gaze and Glow functioned as an exploratory intervention that brought the hidden emotional and cognitive work of editing into public view. By combining experimental video narratives, sensor-based interactions, and persona-driven storytelling, we invited viewers to experience editing not only as a personal, internal activity but as one deeply shaped by imagined and real audiences. In contrast to conventional framings of editing as a purely functional task \cite{bharati_detecting_2016, ying_retouchingffhq_2023}, we approached it as a socially situated and emotionally charged experience.
This framing also aligns with broader discussions of authenticity and visibility among creatives navigating algorithmically mediated platforms. As Cotter argues, users often engage in what she calls the “visibility game,” strategically curating content to align with perceived algorithmic rules \cite{cotter_playing_2019}. Our installation responds to such dynamics by making visible the hidden labor and emotional stakes behind those curated outputs.

A central insight from our design process is the value of interpretive openness in interaction design. Drawing on prior work in embodied interaction and ambiguous design \cite{gaver_ambiguity_2003}, we intentionally avoided prescribing how viewers should interact with the installation. We did not provide explicit instructions about how proximity or group size would affect the display. Instead, viewers discovered these relationships through playful experimentation, co-constructing meaning through interaction and discussion. This ambiguity encouraged diverse interpretations—some users described a sense of vulnerability, while others felt recognized and validated. This variation in affective responses highlights the importance of designing for flexible and personally meaningful engagement, particularly when making private practices public.

\subsubsection*{Rethinking Memory and Forgetting in Editing Practices}
Prior research in HCI has considered the roles of memory and forgetting in social media systems, particularly regarding content persistence and ephemerality \cite{chiu_last_2021, ankenbauer_times_2025, bannon_forgetting_2006, zhao_many_2013}. Platforms like Facebook and Snapchat represent opposing models: one curates memories over time (e.g., “On This Day” \cite{jonathan_gheller_introducing_2015}), while the other encourages impermanence and erasure \cite{xu_automatic_2016}. Yet, as Prey (2018) and others have argued, what is at stake is not only what is remembered and how, but also what is not remembered, and who controls that forgetting \cite{prey_personal_2018}.

Current platforms rarely capture or acknowledge the pre-publication editing process, treating it as ephemeral, storing only fragments in drafts or caches that disappear once a post is published. At the same time, user actions are continuously tracked by the platform’s analytics systems, creating a dissonance between users’ lack of access to their own editing histories and the platform’s persistent surveillance. Sas et al.’s concept of performative forgetting critiques this dynamic \cite{sas_design_2016}, where platforms simulate erasure while retaining control over user data \cite{ankenbauer_times_2025}.

Our findings challenge this binary of deletion vs. archiving \cite{xu_automatic_2016}. We argue for a rethinking of the editing process not as disposable, but as a site of reflection, memory, and agency. Social media platforms could introduce features that allow users to revisit, reflect on, and even optionally share parts of their editing journey. Rather than forcing users to choose between total invisibility and total exposure, systems could support selective visibility \cite{flyverbom_transparency_2016}, where users determine what aspects of their behind-the-scenes processes they want to disclose. This suggestion resonates with recent research by De and Lu, who show how novice poets on Instagram continuously negotiate their artistic intent and authenticity within the constraints of algorithmic visibility \cite{de_poetsofinstagram_2024}. Like our participants, these users adapt their creative labor in response to platform logics, pointing to a broader need for design features that support intentional self-presentation without sacrificing personal integrity and authenticity.

By exposing the invisible and often unacknowledged efforts of social media editing through Gaze and Glow, our work surfaces how these processes are deeply entangled with questions of authenticity, agency, and performativity. We argue that making editing practices visible not only fosters new interpretations of what it means to be “authentic” online, but also opens up possibilities for giving users greater agency in how they manage visibility, memory, and expression in platform-mediated contexts.

\section{Limitations and Future Work} 
We acknowledge the limitations of our study. As Gaze and Glow primarily relied on visual interaction, it presented accessibility barriers for individuals with visual impairments. Future iterations could explore multimodal interaction techniques to foster more inclusive engagement. Additionally, while the installation provoked reflection, it did not directly evaluate the impact of such design interventions within platform interfaces. Future work could involve embedding selective visibility and editing traceability features into social media systems and testing how they support users’ agency and authenticity at a local scale.

\section{Conclusion}
This paper introduced Gaze and Glow, an interactive installation that makes visible the often-overlooked emotional and cognitive efforts embedded in social media editing practices. Through the creation of narrative personas, experimental video storytelling, and sensor-based interaction design, the installation foregrounded the social and performative nature of editing, an act typically experienced in private. By deploying the installation in a public exhibition and analyzing viewer responses, we explored how the exposure of previously invisible editing processes can prompt new interpretations of authenticity and users' agency. Our findings reveal that making editing visible is both affirming and unsettling recognition for users’ hidden efforts. We advocate for future systems that allow users to engage more deliberately with their editing histories, reclaim agency over what is remembered and shared, and embrace the complexities of crafting authenticity in online life.

\bibliographystyle{ACM-Reference-Format}
\bibliography{citation}


\begin{thebibliography}{26}


\ifx \showCODEN    \undefined \def \showCODEN     #1{\unskip}     \fi
\ifx \showDOI      \undefined \def \showDOI       #1{#1}\fi
\ifx \showISBNx    \undefined \def \showISBNx     #1{\unskip}     \fi
\ifx \showISBNxiii \undefined \def \showISBNxiii  #1{\unskip}     \fi
\ifx \showISSN     \undefined \def \showISSN      #1{\unskip}     \fi
\ifx \showLCCN     \undefined \def \showLCCN      #1{\unskip}     \fi
\ifx \shownote     \undefined \def \shownote      #1{#1}          \fi
\ifx \showarticletitle \undefined \def \showarticletitle #1{#1}   \fi
\ifx \showURL      \undefined \def \showURL       {\relax}        \fi
\providecommand\bibfield[2]{#2}
\providecommand\bibinfo[2]{#2}
\providecommand\natexlab[1]{#1}
\providecommand\showeprint[2][]{arXiv:#2}

\bibitem[Ankenbauer and Brewer(2025)]%
        {ankenbauer_times_2025}
\bibfield{author}{\bibinfo{person}{Sam~Addison Ankenbauer} {and} \bibinfo{person}{Robin~N. Brewer}.} \bibinfo{year}{2025}\natexlab{}.
\newblock \showarticletitle{Time's {Sublimest} {Target}: {Practices} of {Forgetting} in {HCI} and {CSCW}}.
\newblock \bibinfo{journal}{\emph{Proc. ACM Hum.-Comput. Interact.}} \bibinfo{volume}{9}, \bibinfo{number}{1} (\bibinfo{date}{Jan.} \bibinfo{year}{2025}), \bibinfo{pages}{GROUP32:1--GROUP32:24}.
\newblock
\urldef\tempurl%
\url{https://doi.org/10.1145/3701211}
\showDOI{\tempurl}


\bibitem[Bannon(2006)]%
        {bannon_forgetting_2006}
\bibfield{author}{\bibinfo{person}{Liam~J. Bannon}.} \bibinfo{year}{2006}\natexlab{}.
\newblock \showarticletitle{Forgetting as a feature, not a bug: the dualityof memory and implications for ubiquitous computing}.
\newblock \bibinfo{journal}{\emph{CoDesign}} \bibinfo{volume}{2}, \bibinfo{number}{1} (\bibinfo{date}{March} \bibinfo{year}{2006}), \bibinfo{pages}{3--15}.
\newblock
\showISSN{1571-0882}
\urldef\tempurl%
\url{https://doi.org/10.1080/15710880600608230}
\showDOI{\tempurl}
\newblock
\shownote{Publisher: Taylor \& Francis \_eprint: https://doi.org/10.1080/15710880600608230}.


\bibitem[Bernstein et~al\mbox{.}(2013)]%
        {bernstein_quantifying_2013}
\bibfield{author}{\bibinfo{person}{Michael~S. Bernstein}, \bibinfo{person}{Eytan Bakshy}, \bibinfo{person}{Moira Burke}, {and} \bibinfo{person}{Brian Karrer}.} \bibinfo{year}{2013}\natexlab{}.
\newblock \showarticletitle{Quantifying the invisible audience in social networks}. In \bibinfo{booktitle}{\emph{Proceedings of the {SIGCHI} {Conference} on {Human} {Factors} in {Computing} {Systems}}} \emph{(\bibinfo{series}{{CHI} '13})}. \bibinfo{publisher}{Association for Computing Machinery}, \bibinfo{address}{New York, NY, USA}, \bibinfo{pages}{21--30}.
\newblock
\showISBNx{978-1-4503-1899-0}
\urldef\tempurl%
\url{https://doi.org/10.1145/2470654.2470658}
\showDOI{\tempurl}


\bibitem[Bharati et~al\mbox{.}(2016)]%
        {bharati_detecting_2016}
\bibfield{author}{\bibinfo{person}{Aparna Bharati}, \bibinfo{person}{Richa Singh}, \bibinfo{person}{Mayank Vatsa}, {and} \bibinfo{person}{Kevin~W. Bowyer}.} \bibinfo{year}{2016}\natexlab{}.
\newblock \showarticletitle{Detecting {Facial} {Retouching} {Using} {Supervised} {Deep} {Learning}}.
\newblock \bibinfo{journal}{\emph{Trans. Info. For. Sec.}} \bibinfo{volume}{11}, \bibinfo{number}{9} (\bibinfo{date}{Sept.} \bibinfo{year}{2016}), \bibinfo{pages}{1903--1913}.
\newblock
\showISSN{1556-6013}
\urldef\tempurl%
\url{https://doi.org/10.1109/TIFS.2016.2561898}
\showDOI{\tempurl}


\bibitem[Braun and Clarke(2020)]%
        {braun_thematic_2020}
\bibfield{author}{\bibinfo{person}{Virginia Braun} {and} \bibinfo{person}{Victoria Clarke}.} \bibinfo{year}{2020}\natexlab{}.
\newblock \showarticletitle{Thematic {Analysis}}.
\newblock In \bibinfo{booktitle}{\emph{Encyclopedia of {Quality} of {Life} and {Well}-{Being} {Research}}}, \bibfield{editor}{\bibinfo{person}{Filomena Maggino}} (Ed.). \bibinfo{publisher}{Springer International Publishing}, \bibinfo{address}{Cham}, \bibinfo{pages}{1--7}.
\newblock
\showISBNx{978-3-319-69909-7}
\urldef\tempurl%
\url{https://doi.org/10.1007/978-3-319-69909-7_3470-2}
\showDOI{\tempurl}


\bibitem[Chiu and Yuan(2021)]%
        {chiu_last_2021}
\bibfield{author}{\bibinfo{person}{Hsuen~Chi Chiu} {and} \bibinfo{person}{Chien Wen~(Tina) Yuan}.} \bibinfo{year}{2021}\natexlab{}.
\newblock \showarticletitle{To {Last} {Long} or to {Fade} {Away}: {Investigating} {Users}' {Instagram} {Post} and {Story} {Practices}}. In \bibinfo{booktitle}{\emph{Companion {Publication} of the 2021 {Conference} on {Computer} {Supported} {Cooperative} {Work} and {Social} {Computing}}} \emph{(\bibinfo{series}{{CSCW} '21})}. \bibinfo{publisher}{Association for Computing Machinery}, \bibinfo{address}{New York, NY, USA}, \bibinfo{pages}{32--35}.
\newblock
\showISBNx{978-1-4503-8479-7}
\urldef\tempurl%
\url{https://doi.org/10.1145/3462204.3481778}
\showDOI{\tempurl}


\bibitem[Cotter(2019)]%
        {cotter_playing_2019}
\bibfield{author}{\bibinfo{person}{Kelley Cotter}.} \bibinfo{year}{2019}\natexlab{}.
\newblock \showarticletitle{Playing the visibility game: How digital influencers and algorithms negotiate influence on Instagram}.
\newblock \bibinfo{journal}{\emph{New Media \& Society}} \bibinfo{volume}{21}, \bibinfo{number}{4} (\bibinfo{year}{2019}), \bibinfo{pages}{895--913}.
\newblock
\urldef\tempurl%
\url{https://doi.org/10.1177/1461444818815684}
\showDOI{\tempurl}
\showeprint{https://doi.org/10.1177/1461444818815684}


\bibitem[De and Lu(2024)]%
        {de_poetsofinstagram_2024}
\bibfield{author}{\bibinfo{person}{Ankolika De} {and} \bibinfo{person}{Zhicong Lu}.} \bibinfo{year}{2024}\natexlab{}.
\newblock \showarticletitle{\#PoetsOfInstagram: Navigating The Practices And Challenges Of Novice Poets On Instagram}. In \bibinfo{booktitle}{\emph{Proceedings of the 2024 CHI Conference on Human Factors in Computing Systems}} (Honolulu, HI, USA) \emph{(\bibinfo{series}{CHI '24})}. \bibinfo{publisher}{Association for Computing Machinery}, \bibinfo{address}{New York, NY, USA}, Article \bibinfo{articleno}{162}, \bibinfo{numpages}{16}~pages.
\newblock
\showISBNx{9798400703300}
\urldef\tempurl%
\url{https://doi.org/10.1145/3613904.3642173}
\showDOI{\tempurl}


\bibitem[Flyverbom(2016)]%
        {flyverbom_transparency_2016}
\bibfield{author}{\bibinfo{person}{Mikkel Flyverbom}.} \bibinfo{year}{2016}\natexlab{}.
\newblock \showarticletitle{Transparency: Mediation and the Management of Visibilities}.
\newblock \bibinfo{journal}{\emph{International Journal of Communication}}  \bibinfo{volume}{10} (\bibinfo{year}{2016}), \bibinfo{pages}{110--122}.
\newblock


\bibitem[Gaver et~al\mbox{.}(1999)]%
        {gaver_design_1999}
\bibfield{author}{\bibinfo{person}{Bill Gaver}, \bibinfo{person}{Tony Dunne}, {and} \bibinfo{person}{Elena Pacenti}.} \bibinfo{year}{1999}\natexlab{}.
\newblock \showarticletitle{Design: {Cultural} probes}.
\newblock \bibinfo{journal}{\emph{Interactions}} \bibinfo{volume}{6}, \bibinfo{number}{1} (\bibinfo{date}{Jan.} \bibinfo{year}{1999}), \bibinfo{pages}{21--29}.
\newblock
\showISSN{1072-5520}
\urldef\tempurl%
\url{https://doi.org/10.1145/291224.291235}
\showDOI{\tempurl}


\bibitem[Gaver et~al\mbox{.}(2003)]%
        {gaver_ambiguity_2003}
\bibfield{author}{\bibinfo{person}{William~W. Gaver}, \bibinfo{person}{Jacob Beaver}, {and} \bibinfo{person}{Steve Benford}.} \bibinfo{year}{2003}\natexlab{}.
\newblock \showarticletitle{Ambiguity as a resource for design}. In \bibinfo{booktitle}{\emph{Proceedings of the {SIGCHI} {Conference} on {Human} {Factors} in {Computing} {Systems}}} \emph{(\bibinfo{series}{{CHI} '03})}. \bibinfo{publisher}{Association for Computing Machinery}, \bibinfo{address}{New York, NY, USA}, \bibinfo{pages}{233--240}.
\newblock
\showISBNx{978-1-58113-630-2}
\urldef\tempurl%
\url{https://doi.org/10.1145/642611.642653}
\showDOI{\tempurl}


\bibitem[Iba et~al\mbox{.}(2017)]%
        {iba_philosophy_2017}
\bibfield{author}{\bibinfo{person}{Takashi Iba}, \bibinfo{person}{Ayaka Yoshikawa}, {and} \bibinfo{person}{Konomi Munakata}.} \bibinfo{year}{2017}\natexlab{}.
\newblock \showarticletitle{Philosophy and methodology of clustering in pattern mining: {Japanese} anthropologist {Jiro} {Kawakita}'s {KJ} method}. In \bibinfo{booktitle}{\emph{Proceedings of the 24th {Conference} on {Pattern} {Languages} of {Programs}}} \emph{(\bibinfo{series}{{PLoP} '17})}. \bibinfo{publisher}{The Hillside Group}, \bibinfo{address}{USA}, \bibinfo{pages}{1--11}.
\newblock
\showISBNx{978-1-941652-06-0}


\bibitem[{Jonathan Gheller}(2015)]%
        {jonathan_gheller_introducing_2015}
\bibfield{author}{\bibinfo{person}{{Jonathan Gheller}}.} \bibinfo{year}{2015}\natexlab{}.
\newblock \bibinfo{title}{Introducing {On} {This} {Day}: {A} {New} {Way} to {Look} {Back} at {Photos} and {Memories} on {Facebook}}.
\newblock
\newblock
\urldef\tempurl%
\url{https://about.fb.com/news/2015/03/introducing-on-this-day-a-new-way-to-look-back-at-photos-and-memories-on-facebook/}
\showURL{%
\tempurl}


\bibitem[Mills et~al\mbox{.}(2018)]%
        {mills_selfie_2018}
\bibfield{author}{\bibinfo{person}{Jennifer~S. Mills}, \bibinfo{person}{Sarah Musto}, \bibinfo{person}{Lindsay Williams}, {and} \bibinfo{person}{Marika Tiggemann}.} \bibinfo{year}{2018}\natexlab{}.
\newblock \showarticletitle{{\textquotedblleft}{Selfie}{\textquotedblright} harm: {Effects} on mood and body image in young women}.
\newblock \bibinfo{journal}{\emph{Body Image}}  \bibinfo{volume}{27} (\bibinfo{date}{Dec.} \bibinfo{year}{2018}), \bibinfo{pages}{86--92}.
\newblock
\showISSN{1740-1445}
\urldef\tempurl%
\url{https://doi.org/10.1016/j.bodyim.2018.08.007}
\showDOI{\tempurl}


\bibitem[Ozimek et~al\mbox{.}(2023)]%
        {ozimek_how_2023}
\bibfield{author}{\bibinfo{person}{Phillip Ozimek}, \bibinfo{person}{Semina Lainas}, \bibinfo{person}{Hans-Werner Bierhoff}, {and} \bibinfo{person}{Elke Rohmann}.} \bibinfo{year}{2023}\natexlab{}.
\newblock \showarticletitle{How photo editing in social media shapes self-perceived attractiveness and self-esteem via self-objectification and physical appearance comparisons}.
\newblock \bibinfo{journal}{\emph{BMC Psychology}} \bibinfo{volume}{11}, \bibinfo{number}{1} (\bibinfo{date}{April} \bibinfo{year}{2023}), \bibinfo{pages}{99}.
\newblock
\showISSN{2050-7283}
\urldef\tempurl%
\url{https://doi.org/10.1186/s40359-023-01143-0}
\showDOI{\tempurl}


\bibitem[Prey and Smit(2018)]%
        {prey_personal_2018}
\bibfield{author}{\bibinfo{person}{Robert Prey} {and} \bibinfo{person}{Rik Smit}.} \bibinfo{year}{2018}\natexlab{}.
\newblock \showarticletitle{From {Personal} to {Personalized} {Memory}: {Social} {Media} as {Mnemotechnology}}.
\newblock In \bibinfo{booktitle}{\emph{A {Networked} {Self} and {Birth}, {Life}, {Death}}}. \bibinfo{publisher}{Routledge}, \bibinfo{address}{New York, NY}.
\newblock
\showISBNx{978-1-315-20212-9}
\newblock
\shownote{Num Pages: 15}.


\bibitem[Reddy and Kumar(2024)]%
        {reddy_teaspoon_2024}
\bibfield{author}{\bibinfo{person}{Ananya Reddy} {and} \bibinfo{person}{Priya~C. Kumar}.} \bibinfo{year}{2024}\natexlab{}.
\newblock \showarticletitle{{\textquoteleft}{A} {Teaspoon} of {Authenticity}{\textquoteright}: {Exploring} {How} {Young} {Adults} {BeReal} on {Social} {Media}}. In \bibinfo{booktitle}{\emph{Proceedings of the 2024 {CHI} {Conference} on {Human} {Factors} in {Computing} {Systems}}} \emph{(\bibinfo{series}{{CHI} '24})}. \bibinfo{publisher}{Association for Computing Machinery}, \bibinfo{address}{New York, NY, USA}, \bibinfo{pages}{1--14}.
\newblock
\showISBNx{9798400703300}
\urldef\tempurl%
\url{https://doi.org/10.1145/3613904.3642690}
\showDOI{\tempurl}


\bibitem[Sas et~al\mbox{.}(2016)]%
        {sas_design_2016}
\bibfield{author}{\bibinfo{person}{Corina Sas}, \bibinfo{person}{Steve Whittaker}, {and} \bibinfo{person}{John Zimmerman}.} \bibinfo{year}{2016}\natexlab{}.
\newblock \showarticletitle{Design for {Rituals} of {Letting} {Go}: {An} {Embodiment} {Perspective} on {Disposal} {Practices} {Informed} by {Grief} {Therapy}}.
\newblock \bibinfo{journal}{\emph{ACM Trans. Comput.-Hum. Interact.}} \bibinfo{volume}{23}, \bibinfo{number}{4} (\bibinfo{date}{Aug.} \bibinfo{year}{2016}), \bibinfo{pages}{21:1--21:37}.
\newblock
\showISSN{1073-0516}
\urldef\tempurl%
\url{https://doi.org/10.1145/2926714}
\showDOI{\tempurl}


\bibitem[Taber and Whittaker(2020)]%
        {taber_finsta_2020}
\bibfield{author}{\bibinfo{person}{Lee Taber} {and} \bibinfo{person}{Steve Whittaker}.} \bibinfo{year}{2020}\natexlab{}.
\newblock \showarticletitle{"{On} {Finsta}, {I} can say '{Hail} {Satan}'": {Being} {Authentic} but {Disagreeable} on {Instagram}}. In \bibinfo{booktitle}{\emph{Proceedings of the 2020 {CHI} {Conference} on {Human} {Factors} in {Computing} {Systems}}} \emph{(\bibinfo{series}{{CHI} '20})}. \bibinfo{publisher}{Association for Computing Machinery}, \bibinfo{address}{New York, NY, USA}, \bibinfo{pages}{1--14}.
\newblock
\showISBNx{978-1-4503-6708-0}
\urldef\tempurl%
\url{https://doi.org/10.1145/3313831.3376182}
\showDOI{\tempurl}


\bibitem[Tiggemann et~al\mbox{.}(2020)]%
        {tiggemann_uploading_2020}
\bibfield{author}{\bibinfo{person}{Marika Tiggemann}, \bibinfo{person}{Isabella Anderberg}, {and} \bibinfo{person}{Zoe Brown}.} \bibinfo{year}{2020}\natexlab{}.
\newblock \showarticletitle{Uploading your best self: {Selfie} editing and body dissatisfaction}.
\newblock \bibinfo{journal}{\emph{Body Image}}  \bibinfo{volume}{33} (\bibinfo{date}{June} \bibinfo{year}{2020}), \bibinfo{pages}{175--182}.
\newblock
\showISSN{1740-1445}
\urldef\tempurl%
\url{https://doi.org/10.1016/j.bodyim.2020.03.002}
\showDOI{\tempurl}


\bibitem[Trieu and Baym(2020)]%
        {trieu_private_2020}
\bibfield{author}{\bibinfo{person}{Penny Trieu} {and} \bibinfo{person}{Nancy~K. Baym}.} \bibinfo{year}{2020}\natexlab{}.
\newblock \showarticletitle{Private {Responses} for {Public} {Sharing}: {Understanding} {Self}-{Presentation} and {Relational} {Maintenance} via {Stories} in {Social} {Media}}. In \bibinfo{booktitle}{\emph{Proceedings of the 2020 {CHI} {Conference} on {Human} {Factors} in {Computing} {Systems}}} \emph{(\bibinfo{series}{{CHI} '20})}. \bibinfo{publisher}{Association for Computing Machinery}, \bibinfo{address}{New York, NY, USA}, \bibinfo{pages}{1--13}.
\newblock
\showISBNx{978-1-4503-6708-0}
\urldef\tempurl%
\url{https://doi.org/10.1145/3313831.3376549}
\showDOI{\tempurl}


\bibitem[Ulman and Stagg(2018)]%
        {ulman_excellences_2018}
\bibfield{author}{\bibinfo{person}{Amalia Ulman} {and} \bibinfo{person}{Natasha Stagg}.} \bibinfo{year}{2018}\natexlab{}.
\newblock \bibinfo{booktitle}{\emph{Excellences \& {Perfections}}}.
\newblock \bibinfo{publisher}{Prestel}, \bibinfo{address}{Munich}.
\newblock
\showISBNx{978-3-7913-8418-4}
\newblock
\shownote{Google-Books-ID: aGo9tAEACAAJ}.


\bibitem[Van~Gelder(1995)]%
        {van_gelder_strange_1995}
\bibfield{author}{\bibinfo{person}{Lindsy Van~Gelder}.} \bibinfo{year}{1995}\natexlab{}.
\newblock \showarticletitle{The strange case of the electronic lover}.
\newblock In \bibinfo{booktitle}{\emph{Computerization and controversy (2nd ed.): value conflicts and social choices}}. \bibinfo{publisher}{Academic Press, Inc.}, \bibinfo{address}{USA}, \bibinfo{pages}{533--546}.
\newblock
\showISBNx{978-0-12-415040-9}


\bibitem[Xu et~al\mbox{.}(2016)]%
        {xu_automatic_2016}
\bibfield{author}{\bibinfo{person}{Bin Xu}, \bibinfo{person}{Pamara Chang}, \bibinfo{person}{Christopher~L Welker}, \bibinfo{person}{Natalya~N. Bazarova}, {and} \bibinfo{person}{Dan Cosley}.} \bibinfo{year}{2016}\natexlab{}.
\newblock \showarticletitle{Automatic {Archiving} versus {Default} {Deletion}: {What} {Snapchat} {Tells} {Us} {About} {Ephemerality} in {Design}}.
\newblock \bibinfo{journal}{\emph{CSCW : proceedings of the Conference on Computer-Supported Cooperative Work. Conference on Computer-Supported Cooperative Work}}  \bibinfo{volume}{2016} (\bibinfo{year}{2016}), \bibinfo{pages}{1662--1675}.
\newblock
\urldef\tempurl%
\url{https://doi.org/10.1145/2818048.2819948}
\showDOI{\tempurl}


\bibitem[Ying et~al\mbox{.}(2023)]%
        {ying_retouchingffhq_2023}
\bibfield{author}{\bibinfo{person}{Qichao Ying}, \bibinfo{person}{Jiaxin Liu}, \bibinfo{person}{Sheng Li}, \bibinfo{person}{Haisheng Xu}, \bibinfo{person}{Zhenxing Qian}, {and} \bibinfo{person}{Xinpeng Zhang}.} \bibinfo{year}{2023}\natexlab{}.
\newblock \showarticletitle{{RetouchingFFHQ}: {A} {Large}-scale {Dataset} for {Fine}-grained {Face} {Retouching} {Detection}}. In \bibinfo{booktitle}{\emph{Proceedings of the 31st {ACM} {International} {Conference} on {Multimedia}}}. \bibinfo{publisher}{ACM}, \bibinfo{address}{Ottawa ON Canada}, \bibinfo{pages}{737--746}.
\newblock
\showISBNx{9798400701085}
\urldef\tempurl%
\url{https://doi.org/10.1145/3581783.3611843}
\showDOI{\tempurl}


\bibitem[Zhao et~al\mbox{.}(2013)]%
        {zhao_many_2013}
\bibfield{author}{\bibinfo{person}{Xuan Zhao}, \bibinfo{person}{Niloufar Salehi}, \bibinfo{person}{Sasha Naranjit}, \bibinfo{person}{Sara Alwaalan}, \bibinfo{person}{Stephen Voida}, {and} \bibinfo{person}{Dan Cosley}.} \bibinfo{year}{2013}\natexlab{}.
\newblock \showarticletitle{The many faces of facebook: experiencing social media as performance, exhibition, and personal archive}. In \bibinfo{booktitle}{\emph{Proceedings of the {SIGCHI} {Conference} on {Human} {Factors} in {Computing} {Systems}}} \emph{(\bibinfo{series}{{CHI} '13})}. \bibinfo{publisher}{Association for Computing Machinery}, \bibinfo{address}{New York, NY, USA}, \bibinfo{pages}{1--10}.
\newblock
\showISBNx{978-1-4503-1899-0}
\urldef\tempurl%
\url{https://doi.org/10.1145/2470654.2470656}
\showDOI{\tempurl}


\end{thebibliography}

\end{document}